\documentclass[aps,prl,twocolumn,showpacs]{revtex4-1}
\usepackage{bm}
\usepackage{mathrsfs}
\usepackage{amsmath}
\usepackage{amssymb}
\usepackage{graphicx}
\usepackage{amsfonts}
\usepackage{amsthm}
\usepackage{color}
\usepackage{dcolumn}
\usepackage{txfonts}

\begin{document}

\title{Quantum criticality at infinite temperature}
\author{Shao-Wen Chen}
\author{Ren-Bao Liu}
\email{rbliu@phy.cuhk.edu.hk}
\homepage{www.phy.cuhk.edu.hk/rbliu}
\affiliation{Department of Physics, Centre for Quantum Coherence, and Institute of Theoretical Physics, The Chinese University of Hong Kong, Hong Kong, China}


\begin{abstract}
Quantum criticality, being important as an indicator of new quantum matters emerging, is known to occur only at zero or low temperature. We find that a quantum probe, if its coherence time is long, can detect quantum criticality at infinitely high temperature. In particular, the echo control over a spin probe can remove the thermal fluctuation effects and hence reveals the quantum fluctuation effects. Probes with quantum coherence time of milliseconds or seconds can be used to study emerging quantum orders that would occur at extremely low temperatures of nano- or pico-Kelvin. This discovery establishes a physical link between time and inverse temperature and provides a new route to the wonderland of quantum matters.
\end{abstract}


\maketitle

Quantum criticality~\cite{Sachdev} accompanies quantum phase transitions at zero temperature, in which the ground state of a macroscopic system changes dramatically at a critical point with tuning a parameter. Quantum criticality is important as it signatures emergence of new quantum matters and new physics~\cite{PhysRevB.65.165113, RevModPhys.77.871, Nature.464.199, RevModPhys.69.1, AP.56.2, RevModPhys.80.885, NaturePhys.2.341}. However, at temperature higher than the system's interaction strength, thermal fluctuations will conceal the quantum criticality. Extremely low temperatures are required for quantum criticality to occur in many interesting systems. For example, for nuclear spins in solids~\cite{RevModPhys.69.1} and cold atoms in optical lattices~\cite{AP.56.2, RevModPhys.80.885, NaturePhys.2.341} temperatures of $10^{-9}$ or even $10^{-12}$ Kelvin are required~\cite{RevModPhys.69.1, PRL.99.120404}. Such limitation excludes many new classes of quantum matters and hence new physics from experimental investigation.
In this Letter, we show that quantum criticality can be observed at infinitely high temperature by measuring the echo signal of a probe spin coupled to a quantum many-body system, because the spin echo can remove the thermal fluctuation effect~\cite{PhysRev.80.580} and therefore reveal the quantum fluctuation effect. We find that quantum criticality that would occur below $10^{-9}$ or even $10^{-12}$ Kelvin can be detected at infinite temperature by a probe spin with coherence time longer than milliseconds or seconds, respectively.

The key is to devise a time-dependent measurement sensitive to the quantum fluctuations. A previous study proposed that quantum criticality can be probed by the Loschmidt echo~\cite{PhysRevLett.96.140604}, which is equivalent to free-induction decay (FID) of a probe spin. Nuclear magnetic resonance experiments that employ pseudo-pure states to simulate effective zero temperature have shown such enhancement of FID for a three-nucleus system~\cite{PhysRevLett.100.100501}. The Loschmidt echo or FID, however, like other conventional measurement of quantum criticality, requires that temperature be much lower than the interaction strength of the system. In magnetic resonance spectroscopy, spin echo can be used to eliminate the effect of thermal fluctuations (or inhomogeneous broadening)~\cite{PhysRev.80.580} with the decay of echo signal induced mostly by quantum fluctuations. Recent study revealed an anomalous decoherence effect due to quantum fluctuations by removing the thermal noise effect~\cite{Zhao_ADE, Huang_ADE}. Thus we are motivated to use spin echo to study quantum criticality at high or even infinite temperature.

We consider the echo signal of a probe spin-$1/2$ coupled to a macroscopic system (a bath). The bath at thermal equilibrium with inverse temperature $\beta$ is described by a density matrix $\rho=e^{-\beta H}/\mathrm{Tr}\big[e^{-\beta H}\big]$. To have conclusive results, we choose an exactly solvable model~\cite{Sachdev}, namely, the one-dimensional Ising model in a transverse field with Hamiltonian
\begin{equation}\label{Hamiltonian}
H_\lambda=-\sum_{j=1}^N\sigma_j^x\sigma_{j+1}^x-\lambda\sum_{j=1}^N\sigma_j^z\equiv H_0+\lambda H_1,
\end{equation}
with periodic boundary condition, where $\sigma_j^{x/y/z}$ is the Pauli matrix of the $j$th spin along the $x/y/z$-axis.
The probe-bath interaction is $g\sigma^z\otimes H_1\equiv \sigma^z\otimes B/2$. The probe strength is chosen to scale with the bath size as $g\sim 1/\sqrt{N}$ which is $\ll 1$ for a large bath so that the bath is only weakly perturbed by the probe. The FID
$L_{\mathrm{FID}}(t)=\left|\mathrm{Tr}\left(e^{-iH_{\lambda+g}t}\rho e^{iH_{\lambda-g}t}\right)\right|$.
In spin echo, the probe spin is flipped ($|\uparrow \rangle \Leftrightarrow |\downarrow\rangle$) at a time $t/2$,
and the spin coherence is measured at $t$. The echo signal $L_{\mathrm{SE}}(t)=\left|\mathrm{Tr}\left(e^{-iH_{\lambda-g}t/2}e^{-iH_{\lambda+g}t/2}\rho e^{iH_{\lambda-g}t/2}e^{iH_{\lambda+g}t/2}\right)\right|$. The spin chain model has no phase transition at finite temperature but has a quantum criticality between a ferromagnetic order for $\lambda<1$ and a paramagnetic order for $\lambda>1$~\cite{Sachdev}. This model has been used previously to demonstrate the effect of quantum criticality on FID~\cite{PhysRevLett.96.140604}. A previous study on spin echo for this model~\cite{epjd_e2007-00266-6}, however, missed the quantum criticality features at high temperature due to inadequate approximation.

\begin{figure}[t]
\includegraphics[width=\columnwidth]{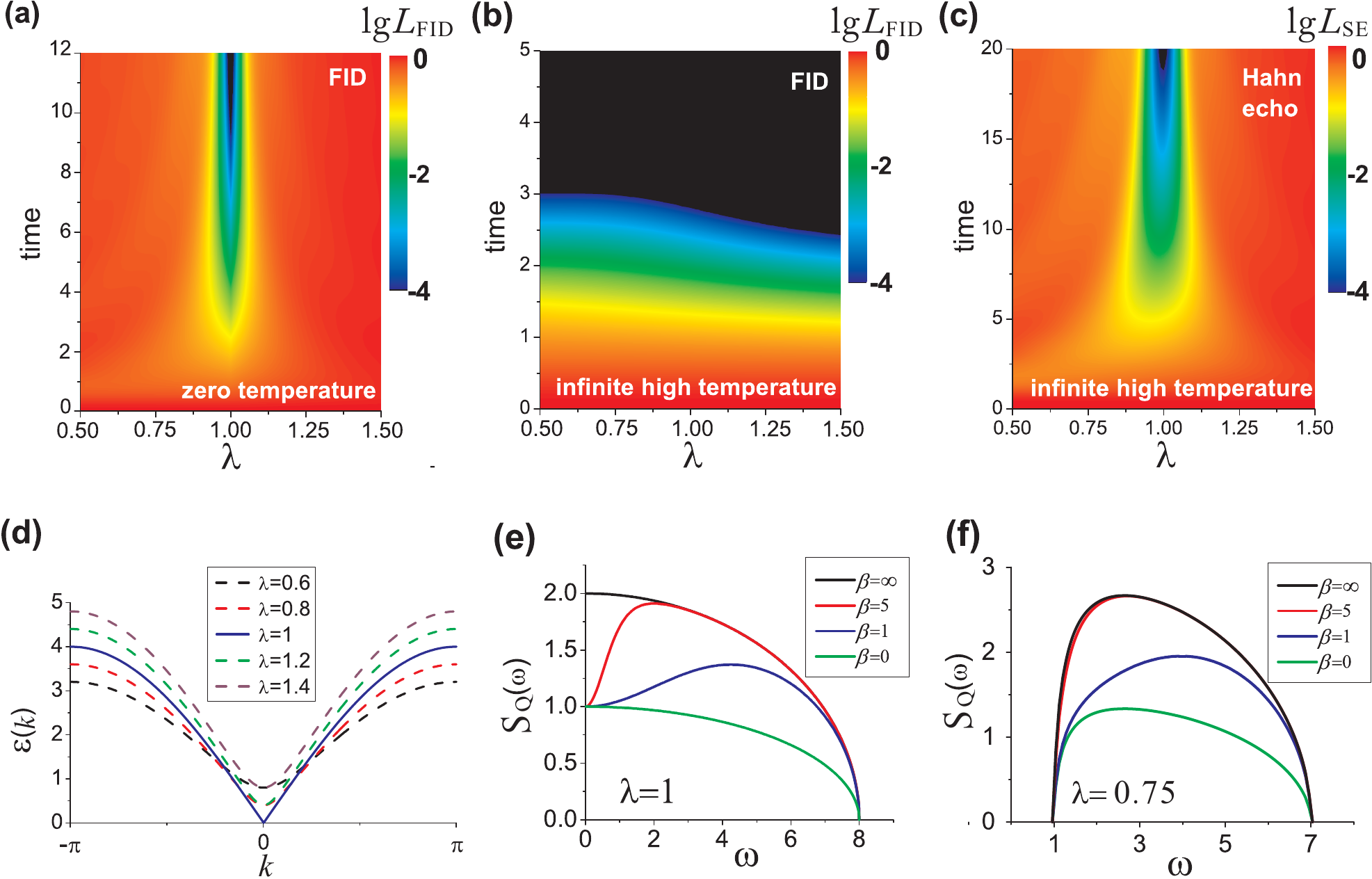}
\caption{\label{Fig1} (Color online) (a) FID of the probe spin versus time and the external field strength $\lambda$ for the bath at zero temperature. (b) The same as (a) but for the bath at infinite temperature. (c) Echo signal of the probe spin versus time and the external field strength for the bath at infinite temperature. (d) Dispersion of elementary excitations in the bath for various external field strength. (e) Noise spectra of quantum fluctuations at various temperatures (indicated by the inverse temperature $\beta$) for external field $\lambda=1$. (f)The same as (e) but with $\lambda=0.75$. The number of spins in the bath is $N=10,000$ and the probe-bath coupling $g=N^{-1/2}=0.01$.}
\end{figure}

Figure~\ref{Fig1}(a) shows that FID of the probe spin is greatly enhanced at the quantum critical point when the temperature is zero ($\beta=\infty$), which is consistent with previous study~\cite{PhysRevLett.96.140604}. The sharp dip at the critical point, however, is blurred with increasing temperature and disappears at infinite temperature [Fig.~\ref{Fig1}(b)]. In contrast, the spin echo signal [Fig.~\ref{Fig1}(c)] presents enhanced decoherence at the critical point even at infinite temperature ($\beta=0$). Both in FID and spin echo, the critical feature is pronounced only when $t\gg 1$.

The above-mentioned phenomena can be understood from the noise spectrum of the bath. The fluctuation of the local field $B\equiv 2gH_1$ has both thermal and quantum components, with the correlation function $C(t)=\langle \tilde{B}(t)\tilde{B}(0)\rangle-\langle \tilde{B}(t)\rangle\langle\tilde{B}(0)\rangle$, where $\langle O\rangle\equiv \text{Tr}\left[\rho O\right]$ and $\tilde{O}(t)\equiv e^{iHt}O e^{-iHt}$. The probe spin decoherence is determined by the noise spectrum $S(\omega)\equiv\int C(t)\exp(i\omega t)dt$. The thermal fluctuation part $S_{\mathrm{th}}=\sum_n P_n\langle n,\lambda|B|n,\lambda\rangle^2-\langle B\rangle^2$ is due to the fact that at finite temperature the bath has a probability distribution $P_n$ in different eigenstates $|n,\lambda\rangle$ that yield different local fields. At zero temperature the thermal fluctuation vanishes. In general, the local field operator $H_1$ does not commute with the bath interaction Hamiltonian $H_0$. Thus transitions between different eigenstates by elementary excitations lead to quantum fluctuation.  The quantum fluctuation is dynamical and has a spectrum $S_Q(\omega)=2\pi\sum_{n\neq m}\delta(\omega-E_n+E_m)P_n\langle n,\lambda|B|m,\lambda\rangle\langle m,\lambda|B|n,\lambda\rangle$.
At high temperature, the thermal fluctuation is usually much stronger than the quantum fluctuation. As the thermal fluctuation is static, its effect on the probe spin decoherence can be removed by spin echo~\cite{PhysRev.80.580}. Then the decoherence is determined by the dynamical quantum fluctuation. In the long time limit, the decoherence would be mostly due to the low-frequency noise caused by low-energy or long-wavelength excitations in the bath, which are particularly important in quantum criticality. The excitation energy as a function of wavevector is $\varepsilon(k)=2\sqrt{1-2\lambda\cos k+\lambda^2}$.
 The excitation has a finite energy gap except for the critical point $\lambda=1$ [Fig.~\ref{Fig1}(d)]. The quantum fluctuation spectrum is gapless at the critical point [Fig.~\ref{Fig1}(e)] and has a low-frequency cut-off for $\lambda\neq 1$ [Fig.~\ref{Fig1}(f)]. Gapless fluctuation emerging at the critical point is responsible for the decoherence enhancement in the long time limit.

\begin{figure}[t]
\includegraphics[width=0.9\columnwidth]{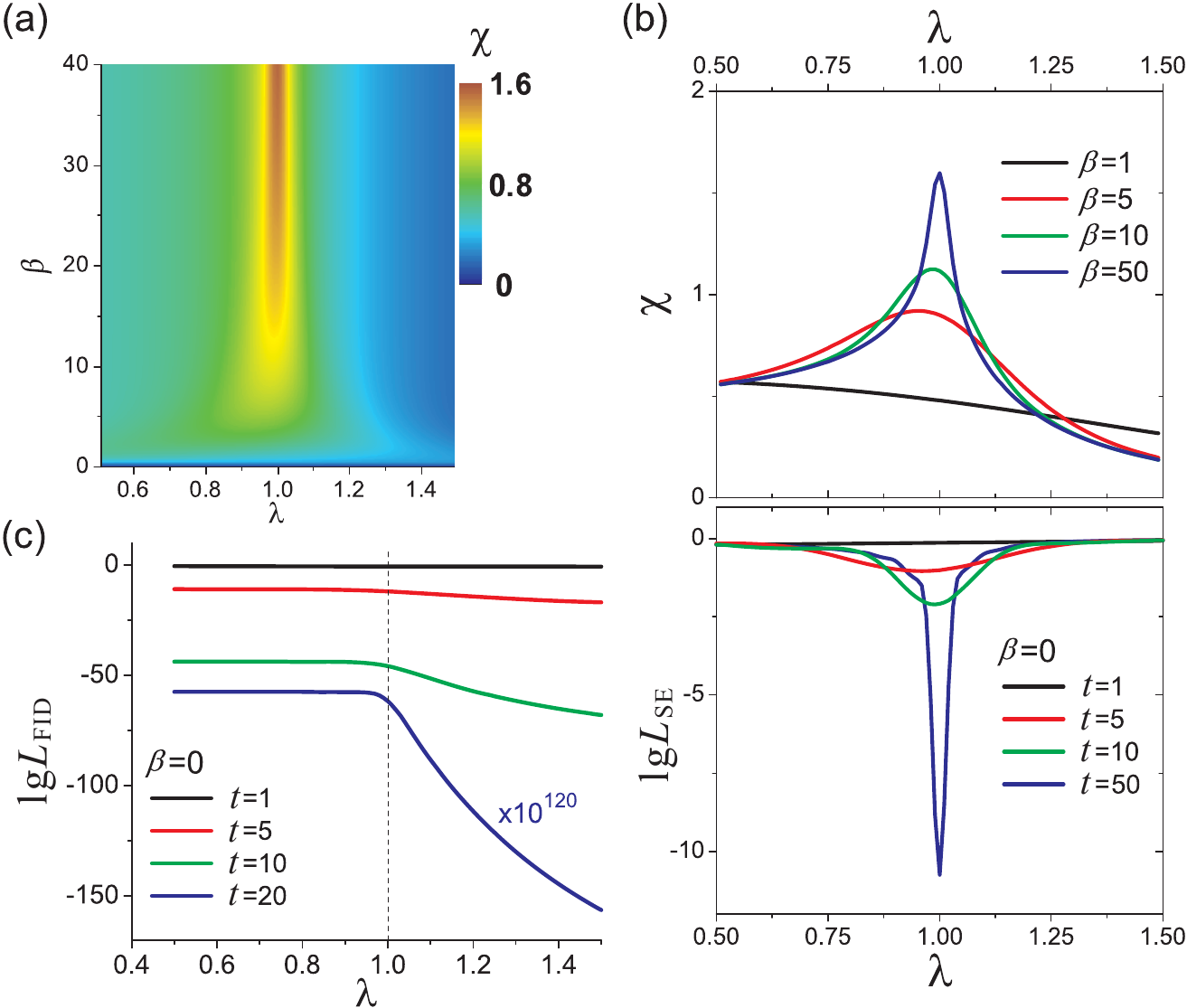}
\caption{\label{Fig2}  (Color online)  (a) Susceptibility of the spin chain bath as a function of the external field $\lambda$ and inverse temperature $\beta$. (b) Susceptibility of the bath (upper panel) and probe spin echo signals (lower panel) as functions of the external field $\lambda$ for various inverse temperature $\beta$ and echo time $t$, respectively. (c) Free-induction decay of the probe spin as a function of the external field for various time $t$. The coherence at $t=20$ is amplified by $10^{120}$.}
\end{figure}

We further explore the correspondence between the time and inverse temperature required for the quantum criticality to pronounce. Fig.~\ref{Fig1}(c) shows that the decoherence enhancement at the critical point is visible only at large $t$. Fig.~\ref{Fig2}(a) presents the magnetic susceptibility of the spin chain $\chi\equiv -N^{-1}\frac{\partial}{\partial\lambda}\sum_i\langle\sigma^z_i\rangle$ versus the inverse temperature $\beta$ and the external field strength $\lambda$ in the thermodynamic limit ($N\rightarrow\infty$). The quantum criticality feature is visible for large $\beta$ (e.g., $>10$), resembling the echo signal as a function of time and field strength in Fig.~\ref{Fig1}(c). Fig.~\ref{Fig2}(b) shows clearly that the sharp features at the critical point are pronounced at similar values of $\beta$ and $t$ in the susceptibility and probe spin coherence echo, respectively.
Actually, if the measurement time is long enough, even the FID would display a sudden transition at the critical point [Fig.~\ref{Fig2}(c)], which, however, is far beyond feasible measurement since the remaining coherence in FID is as little as $10^{-180}$ (at $t=20$).

The probe spin decoherence and the bath susceptibility are intrinsically related as both of them are rooted in the local field fluctuations. Under the weak-probe condition, the probe spin decoherence is determined by the noise correlation function $C(t_1-t_2)$ as
\begin{equation}\label{Lspec}
\ln [{L_\alpha }(t)] =  - \frac{1}{2}\int_0^t {\int_0^t {d{t_1}d{t_2}C\left( {{t_1} - {t_2}} \right){f_\alpha }({t_1}){f_\alpha }({t_2})} } ,
\end{equation}
with $\alpha=$ FID or SE corresponding to the FID and the spin echo signal, respectively. The modulation functions $f_{\mathrm{FID}}(t')=1$ and $f_{\mathrm{SE}}(t')=1$ for $t'\in [0,t/2]$ and $f_{\mathrm{SE}}(t')=-1$ for $t'\in [t/2,t]$. The magnetic susceptibility is determined by the correlation function as
\begin{equation}\label{chi}
\chi  = \frac{1}{{4N{g^2}}}\int_0^\beta  {d\tau C\left( {i\tau } \right)} .
\end{equation}
The susceptibility has the form of the noise correlation function integrated along the imaginary axis in the complex plane of the time. Therefore for the quantum criticality to be pronounced, the inverse temperature required in susceptibility measurement should be in the same order of the time duration required in probe coherence measurement.

\begin{figure}[t]
\includegraphics[width=\columnwidth]{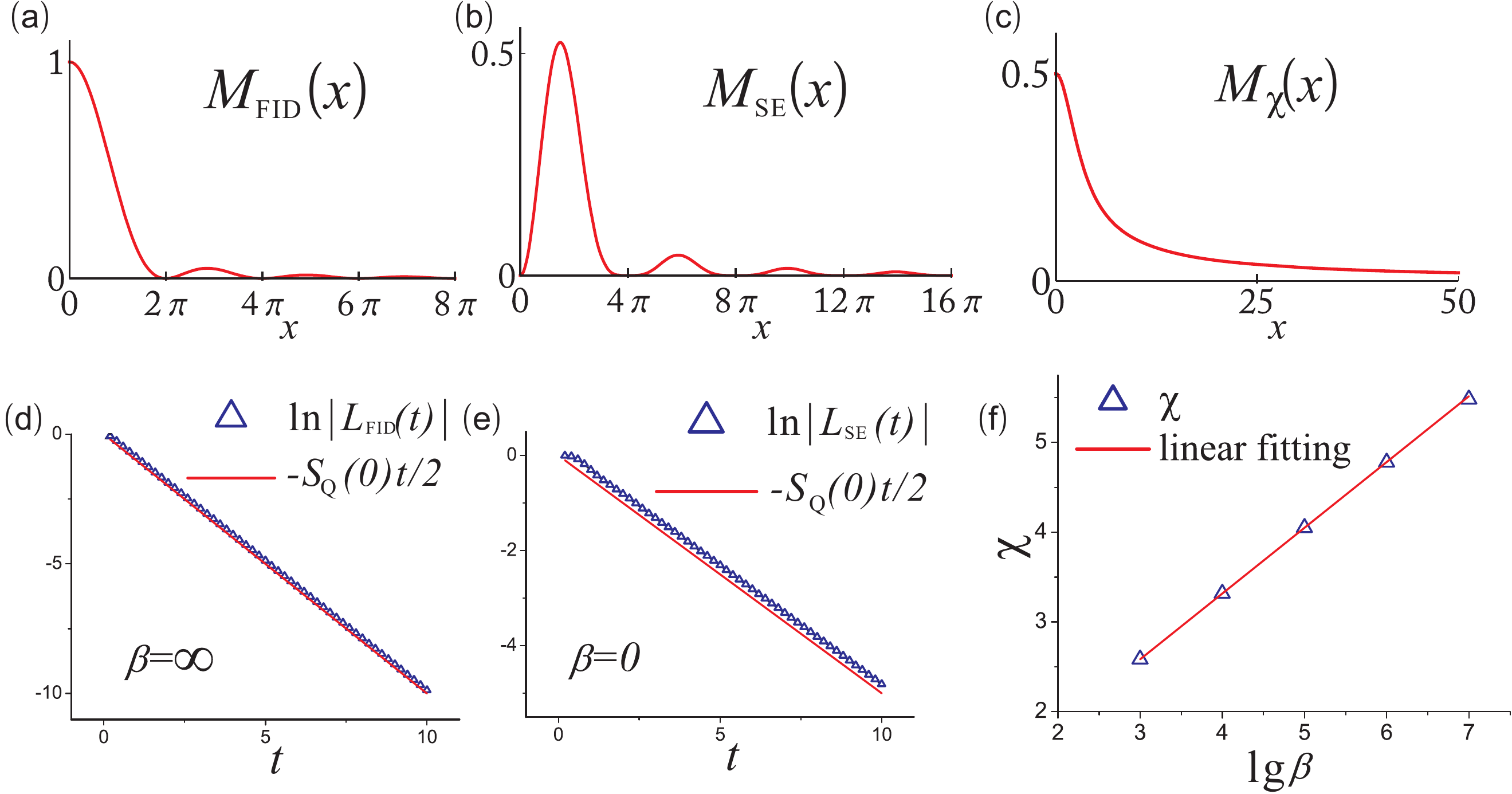}
\caption{\label{Fig3}  (Color online)  (a), (b) \& (c) show in turn the filter functions for free-induction decay, spin echo and magnetic susceptibility as functions of the scaled frequency ($x = \omega t$ or $x = \omega \beta$). (d) The symbols are $\mathrm{ln}|L_\mathrm{FID}|$ as a function of time at $\lambda=1$  and the solid line is the linear fitting. (e) The same as (d) but for the spin echo signal. (f) The symbols are the magnetic susceptibility as a function of $\mathrm{lg}(\beta)$ at $\lambda=1$ and the line is the linear fitting. }
\end{figure}

  By Fourier transform of Eq.~(\ref{Lspec}), the probe spin decoherence is determined by the noise spectrum as~\cite{PhysRevB.77.174509}.
 \begin{subequations}
\begin{align}
\ln \left[ {{L_{{\rm{FID}}}}(t)} \right] =&  - {S_{{\rm{th}}}}{t^2} - t\int_0^\infty  {\frac{{dx}}{{2\pi }}{S_Q}\left( x /t\right){M_{{\rm{FID}}}}\left( x \right)} , \\
 \ln \left[ {{L_{{\rm{SE}}}}(t)} \right] =&  - t\int_0^\infty  {\frac{{dx}}{{2\pi }}{S_Q}\left( x/t\right){M_{{\rm{SE}}}}\left( x \right)},
 \end{align}
 \label{Lspec2}
 \end{subequations}
 where the filter functions $M_{\mathrm{FID}}(x)=\mathrm{sinc}^2 (x/2)$ and $M_{\mathrm{SE}}(x)=\frac{1}{16}x^2\mathrm{sinc}^4 (x/4)$ are determined by Fourier transform of the modulation functions $f_{\mathrm{FID}}(t)$ and $f_{\mathrm{SE}}(t)$, respectively. Similarly, the magnetic susceptibility of the bath is
\begin{equation}\label{chispec}
\chi \left( \beta  \right) = \frac{\beta }{2}{S_{{\rm{th}}}} + \int_0^\infty  {{S_Q}\left( x/\beta  \right){M_\chi }\left( x \right)\frac{{dx}}{{2\pi }}} ,
\end{equation}
with ${M_\chi }\left( x \right) = {x^{ - 1}}\tanh \left( x \right)$ being the corresponding filter function derived by Fourier transform of Eq.~(\ref{chi}). Above the frequency has been scaled by $x=\omega t$ or $\omega\beta$. The modulation functions are plot in Fig.~\ref{Fig3}(a)-(c).

Now we study the critical behaviors at the critical point. The probe spin FID at the critical point diverges linearly with time at zero temperature ($\beta=\infty$) [Fig.~\ref{Fig3}(d)]. In the long time limit ($t\gg 1$), the probe spin decoherence is determined by the low-frequency noise, so in Eq.~(\ref{Lspec2}) the quantum noise spectrum can be approximated as a constant $S_Q(0)$.  Also at zero temperature, the thermal noise is zero. Therefore the FID at the critical point scales with time by ${\rm{ln}}\left| {{L_{{\rm{FID}}}}} \right| \approx  - t{S_Q}(0)\int_0^\infty  {\frac{{dx}}{{2\pi }}} {M_{{\rm{FID}}}}\left( x \right) =  - \frac{1}{2}t{S_Q}(0)$, as observed in Fig.~\ref{Fig3}(d).  The spin echo at infinite temperature ($\beta=0$) also scales linearly with time at the critical point. The filter function of spin echo [see Fig.~\ref{Fig3}(b)] is zero at zero frequency and has its maximum at ${\omega _0} \approx 4.7{t^{ - 1}}$. Similar to the FID case, the low-frequency noise dominates the spin echo decay and leads to the critical scaling ${\rm{ln}}|{L_{{\rm{SE}}}}| \sim  - \frac{1}{2}t{S_{\rm{Q}}}\left( 0 \right)$, which is confirmed in Fig.~\ref{Fig3}(e). In the case of magnetic susceptibility, the modulation function approaches to $1/x$ as $x=\omega\beta\rightarrow\infty$. As the noise spectrum has a high-frequency cutoff [see Fig.~\ref{Fig1}(e)], the integration in Eq.~(\ref{chispec}) leads to the logarithm divergence with the inverse temperature, i.e., $\chi \sim \int^{4\left| {\lambda  + {\lambda _c}} \right|\beta } {{x^{ - 1}}dx} \sim \lg \beta  $, as observed in Fig.~\ref{Fig3}(f).

\begin{figure}[t]
\includegraphics[width=0.9\columnwidth]{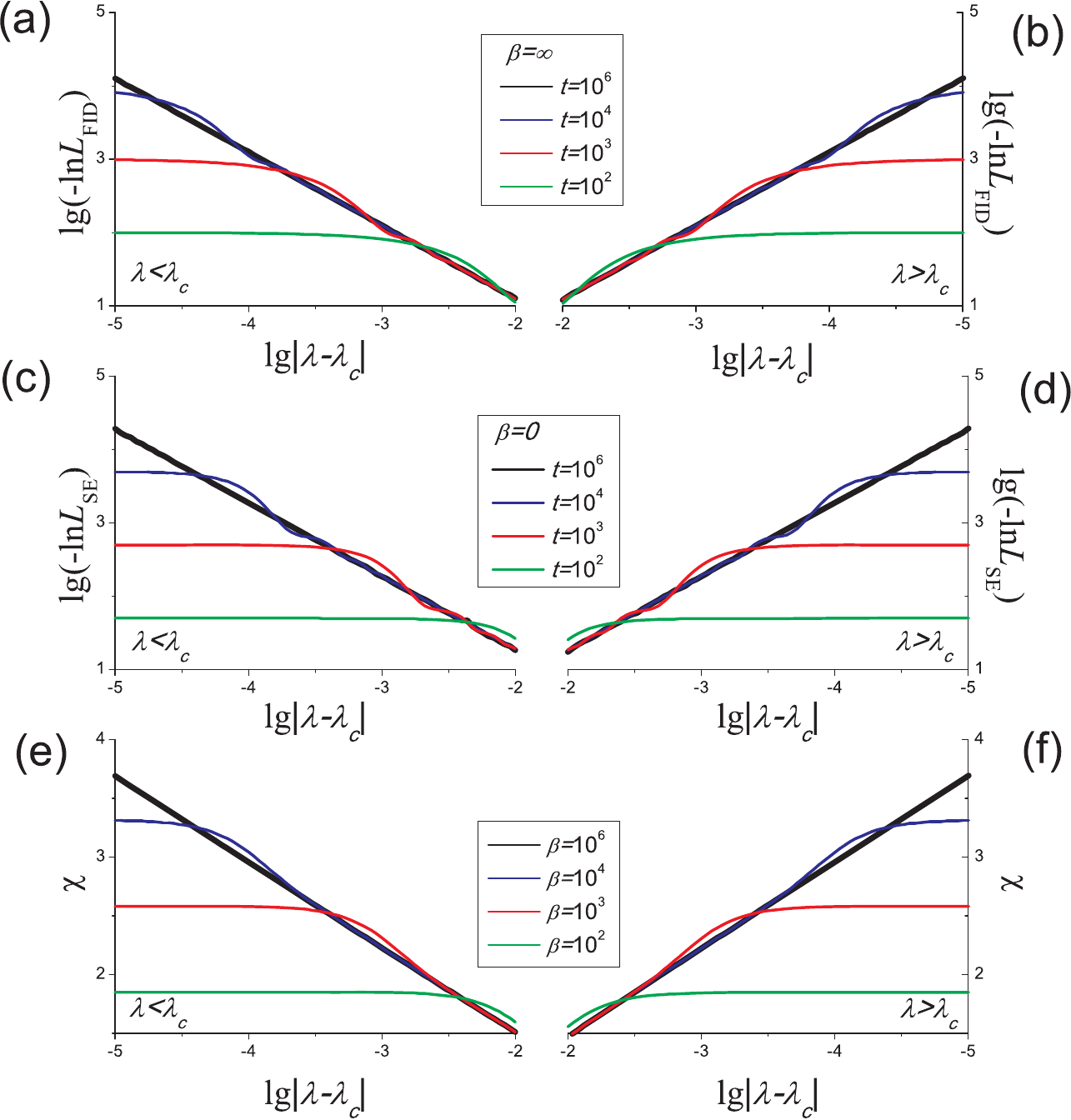}
\caption{\label{Fig4}  (Color online)   (a, b) Free-induction decay of the probe spin as a function of $\mathrm{lg}|\lambda-\lambda_c|$ for (a) $\lambda<\lambda_c$ and (b) $\lambda>\lambda_c$ at various evolution time. (c, d) the same as (a, b), but for the spin echo. (e, f) Magnetic susceptibility as a function of $\mathrm{lg}|\lambda-\lambda_c|$ for (e) $\lambda<\lambda_c$ and (f) $\lambda>\lambda_c$ at various temperature.}
\end{figure}

Figure~\ref{Fig4} shows the respective scaling of the probe spin coherence or the susceptibility with the external field approaching the critical point ($\left| {\lambda  - {\lambda _{\rm{c}}}} \right| \to 0$) for large evolution time or low temperature. To deduce the critical behavior at zero-temperature or infinite time ($\beta,t=\infty$), we choose the temperature or time such that $\left| {\lambda  - {\lambda _c}} \right|\beta  \gg 1$ or $\left| {\lambda  - {\lambda _c}} \right|t  \gg 1$. For a large time, the scaled noise spectrum $S_Q(x/t)$ is spanned in the range from the low-frequency cut-off $4|\lambda-\lambda_c|t$ to the high-frequency cut-off $4|\lambda+\lambda_c|t$. With the condition $\left| {\lambda  - {\lambda _c}} \right|t \gg 1$, the filter functions  for FID or spin echo decay with the scaled frequency as $x^{-2}$. Therefore the probe spin decoherence in FID or spin echo, up to a structure factor in the order of one,
$ - {\rm{ln}}\left| {{L_{{\rm{FID/SE}}}}} \right| \sim \int_{4\left| {\lambda  - {\lambda _c}} \right|t}^{4\left| {\lambda  + {\lambda _c}} \right|t} {{x^{ - 2}}dx}  \sim {\left| {\lambda  - {\lambda _c}} \right|^{ - 1}}$,
with an inverse linear divergence. Note that the divergence is determined by the low-frequency cut-off due to the excitation gap in the bath. Such a critical divergence is shown in Fig.~\ref{Fig4}(a)-(d). The oscillation features in the decoherence are due to the oscillations in the filter functions [see Fig.~\ref{Fig3}(a) and (b)]. The inverse linear scaling is violated when the long time condition $\left| {\lambda  - {\lambda _c}} \right|t \gg 1$ is not fulfilled (since the filter functions converge to a constant rather than diverge as $x^{-2}$ at the zero frequency).  The critical scaling of the susceptibility can be analyzed similarly. Now that the modulation function decays with frequency as $1/x$, the scaling relation becomes
$\chi  \sim \int_{4\left| {\lambda  - {\lambda _c}} \right|t}^{4\left| {\lambda  + {\lambda _c}} \right|t} {{x^{ - 1}}dx}  \sim \ln \left| {\lambda  - {\lambda _c}} \right|$, with a logarithm divergence, as shown in Fig.~\ref{Fig4}(e) and (f).

As demonstrated in Fig.~\ref{Fig3} and Fig.~\ref{Fig4}, the probe coherence is actually more sensitive to the criticality than the conventional susceptibility (linear versus logarithm divergence). This is due to the fact that the filter functions for the FID and spin echo decay faster with increasing the frequency and therefore are more sensitive to the low-frequency dynamics, which is responsible for the critical phenomena.

 In summary, we propose a new route to the wonderland of quantum matters, in lieu of lowering temperature to the critical regime. The time-inverse temperature correspondence enables utilization of long coherence time to detect at high temperature quantum criticality and hence new quantum matters which would occur at extremely low temperature.
 A wide range of quantum probes, such as cold atoms in optical lattices or traps, defect spins in diamond, donor spins in silicon, and nuclear spins, have coherence time from milliseconds to seconds~\cite{PhysRevLett.106.240801, NatureMaterials.8.383, NatureMaterials.2011, PhysRevB.71.014401}, and therefore can be used to study physics that would otherwise emerge at nano-Kelvin to pico-Kelvin.  Spin echo, by largely removing the thermal fluctuation effect, can prolong the coherence time of a quantum probe. It is conceivable that longer coherence time and therefore richer physics can be brought into the reach by applying many-pulse dynamical decoupling control over the probe~\cite{PhysRevA.77.052112, Nature.461.1265, Science.330.60, PhysRevLett.105.200402}. Therefore it is envisaged that dynamical decoupling become a useful tool to study many-body correlations in baths, beyond its existing applications in noise spectrum measurement~\cite{NaturePhys.7.565} and high-sensitivity metrology~\cite{NatureNano.6.242}.

\begin{acknowledgments}
We thank Zhan-Feng Jiang, Nan Zhao, Sen Yang, J. Wrachtrup, C. P. Sun, J. Du, X. H. Peng, and L. J. Sham for discussions. This work was supported by Hong Kong RGC/GRF CUHK402208 \& CUHK402410, The CUHK Focused Investments Scheme, NSFC No.11028510, and Hong Kong RGC/CRF HKU8/CRF/11G.
\end{acknowledgments}

%

\end{document}